\documentclass[traditabstract]{aa}  

\usepackage{graphicx}
\usepackage{txfonts}
\usepackage{longtable}

\usepackage{natbib}\usepackage{hyperref}
\hypersetup{
    colorlinks=true,
    linkcolor=blue,
    filecolor=magenta,      
    urlcolor=cyan,
    citecolor=blue
}

\usepackage{url}
\usepackage{float,color}
\usepackage[figuresright]{rotating}
\usepackage[free-standing-units]{siunitx}
\usepackage{xcolor}
\usepackage{soul}
\bibpunct{(}{)}{;}{a}{}{,}

\def\Msun{\,{\rm M}_\odot}

\def\teff{T_{\rm eff}}

\def\logg{\log({\rm g})}

\def\star{{\rm star}}

\newcommand{\msun}{{\rm M_\odot}}

\begin{document} 

\title{The solar abundance problem and eMSTOs in clusters}
\subtitle{Can they be explained by the accretion of protoplanetary discs during stellar evolution?}

\author{Richard Hoppe\inst{1,2} \and 
       Maria Bergemann\inst{1} \and 
       Bertram Bitsch\inst{1} \and 
       Aldo Serenelli\inst{3, 4}}
       
\institute{Max Planck Institute for Astronomy, 69117, Heidelberg, Germany \email{hoppe@mpia-hd.mpg.de}
\label{inst1}
\and
Ruprecht-Karls-Universit\"at, Grabengasse 1, 69117 Heidelberg, Germany 
\label{inst2}
\and
Institute of Space Sciences (ICE, CSIC), Carrer de Can Magrans S/N, E-08193, Cerdanyola del Valles, Spain
    \and
Institut d'Estudis Espacials de Catalunya (IEEC), Carrer Gran Capita 2, E-08034, Barcelona, Spain
\label{inst3}}

\date{Received xxx / Accepted yyy}

\abstract{We study the impact of accretion from protoplanetary discs on stellar evolution of AFG-type stars. We use a simplified disc model computed using the Two-Pop-Py code that contains the growth and drift of dust particles in the protoplanetary disc. It is used to model the accretion scenarios for a range of physical conditions of protoplanetary discs.  Two limiting cases are combined with the evolution of stellar convective envelopes computed using the \texttt{Garstec} stellar evolution code. We find that the accretion of metal-poor (gas) or metal-rich (dust) material has a significant impact on the chemical composition of the stellar convective envelope. As a consequence, the evolutionary track of the star diverts from the standard scenario predicted by canonical stellar evolution models, which assume a constant and homogeneous chemical composition after the assembly of the star has finished. In the case of the Sun, we find a modest impact on the solar chemical composition. Accretion of metal-poor material indeed reduces the overall metallicity of the solar atmosphere, and it is consistent, within the uncertainty, with the solar $Z$ reported by Caffau et al. (2011), but our model is not consistent with the measurement by Asplund et al. (2009). Another effect is the change of the position of the star in the colour-magnitude diagram. We compare our predictions to a set of open clusters from the Gaia DR2 and show that it is possible to produce a scatter close to the TO of young clusters that could contribute to explain the observed scatter in CMDs. Detailed measurements of metallicities and abundances in the nearby open clusters will provide a stringent observational test of our proposed scenario.}

\titlerunning{discs and stellar evolution}

\authorrunning{Hoppe et al.}

\maketitle
\section{Introduction}
Over the past decades, major progress has been made in studies of proto-planetary (hereafter, PP) discs through modelling \citep[e.g]{Dullemond2002, dullemond2004global, oka2011evolution, birnstiel2011dust, birnstiel2012simple, bitsch2013stellar, baillie2015time} and observations \cite[e.g.][]{ALMA2015}. It has been recognized that discs are ubiquitous around young stars and a significant fraction of those has been observed with facilities like ALMA \citep[e.g.][]{clarke2018high, Fedele2018,favre2019gas, van2019alma, diep2019protoplanetary, nazari2019revealing}, Gemini \citep[e.g.][]{salyk2019high}, VLA \citep[e.g.][]{Macias2018}, and SPHERE \citep[e.g.][]{VanBoekel2017, Avenhaus2017, Avenhaus2018}. However, the relationships between disc formation and the presence or absence of planets are still poorly understood. Although gaps and rings detected with large surveys, like ALMA, are thought to be associated with planet formation, the system PDS 70 still remains the only  confirmed detection of a planet in a PP disc \citep{Metchev2004,Hashimoto2012,Keppler2018}. Nonetheless, surveys for planets orbiting solar-like stars demonstrated that planets are common and possibly even outnumber stars in the Milky Way galaxy \cite[e.g.][]{Mayor2011, Cassan2012, Mulders2018}.

The formation of PP discs has an important, yet, so far, unexplored, effect on the formation and evolution of their central stars. Canonical 1D stellar structure models evolve stars from an initially bright and expanded  object along the Hayashi track to the asymptotic giant branch (AGB) or white dwarf (WD) phase, yet assuming that stellar evolution proceeds in isolation \citep[e.g.][]{Pietrinferni2004,Weiss2008,Paxton2011}. However, the population statistics of exo-planets and discs clearly shows that this assumption is not justified. One important aspect of the coupled star-planet evolution is the accretion of the dust and gas from the disc onto the central star. This effect could be especially important during the early phases of stellar evolution, from a $\sim$few to $\sim$10 Myr, especially for $M \ga 1\Msun$ when the star's convective envelope evolves rapidly. In a recent study \citet{Kunitomo2018} highlighted the substantial effect that the accretion from the PP disc has on the pre-MS star. However, they focused on the accretion of metal-poor material, assuming that refractory elements are locked up in planets or planetesimals. Metal-rich accretion in the form of planet engulfment was studied by \citet{Tognelli2016}. \citet{Serenelli2011} explored the effects of accretion onto the early Sun. However, the study employed arbitrary $\dot{M}$-composition scenarios, and no attempt was made to probe different regimes of stellar parameter space (metallicities, ages) other than the Sun. Other than that, strong evidence for the accretion of planetesimals orbiting about WDs has been established \citep[e.g.][]{Vanderburg2015,Gaensicke2019}. Also the differences between abundance- condensation temperature trends of the Sun and sun-like stars can potentially be explained by the presence of planets or accretion of planetesimals  \citep[e.g.][]{Melendez2009,Bedell2018}.
In this study, we model the evolution of stars of 1 to 3 solar masses, to explore the influence of accretion from the PP disc on the metallicity in stellar convective envelopes and on the star's effective temperature and metallicity respectively. The work is motivated by two observational findings. One of them is the discrepancy between the predictions of Standard Solar Model (SSM) for the photospheric solar abundances and helioseismic predictions \citep[e.g.][]{Serenelli2009}. The other is the putative spreads in the colour-magnitude diagrams (CMD) of young open clusters, especially at their turnoff point \citep[e.g.][]{Marino2018, Cordoni2018}.
We employ the Two-Pop-Py (TPP) code to model the evolution of the protoplanetary disc \citep{birnstiel2010} and the \texttt{Garstec} code \citep{Weiss2008} for the evolution of the star. Different from \citet{Kunitomo2018}, we assume any prior on the accretion rate and the chemical composition of the accreted material, but rather employ a diversity of physically-motivated $\dot{M}$-composition scenarios. In this way, we embrace all realistic cases of accretion scenarios that encompass fast early and slow late accretion. The mass of the accreted material is constrained by observational studies \cite[e.g.][]{Beckwith1990, Ballering2019}. The model we employ to analyse these scenarios is simplified, and it does not capture all physical processes in the discs, such as hydrodynamic turbulence, multi-phase structure, detailed chemical composition, and ionization by cosmic rays. However, modelling these processes is beyond the scope of this paper, as we do not attempt to make quantitative statements on the physical properties of the PP discs, but rather focus on the effect of differential accretion of dust and gas on the chemical evolution of the convective envelope as a function of fundamental stellar parameters.
We organise this paper as follows. In Sect. \ref{sec:tpp}, we outline the properties of the protoplanetary disc models. We describe the treatment of the mass accretion rate and its chemical composition in Sect. \ref{sec:chem} and discuss the accretion histories in Sect. \ref{sec:acchis}. The details of stellar evolution models are provided in Sect. \ref{sec:garstec} and their chemical evolution through accretion is explained in Sect. \ref{sec:chemconv}. The results are presented in Sect. \ref{sec:res}. The influence of accretion on the surface composition of the Sun is discussed in Sec. \ref{sec:sun} and we move on with the analysis of the effects of accretion on the physical parameters of stars and observable properties of stellar populations in Sec. \ref{sec:clust}. The conclusions are drawn in Sect. \ref{sec:conclusions}.

\section{Methods}

\subsection{Protoplanetary disc models}\label{sec:tpp}

We employ the "Two-Pop-Py" code \citep[][hereafter, TPP]{birnstiel2012simple} to  derive quantitative estimates of the properties of accretion rate onto the star and, thus, to constrain the range of plausible accretion scenarios. This is a simplified and  computationally inexpensive model, constructed to reproduce the radial evolution of dust surface densities computed by full-fledged protoplanetary disc models \citep{birnstiel2010, Okuzumi2011}. The model includes fragmentation, cratering, and radial transport mechanisms as the factors that limit the dust growth \citep{DominikDullemond2008, Brauer2008}. The material can exist in the dust phase and gas phases, but transition between these physical phases of matter has not been included in our model.

\begin{table}
\centering
  \caption{Input parameters of TPP simulation. The range of $\alpha$ and $\nu_{\rm frag}$ is further discussed in Sect. \ref{sec:acchis}. Note that $M$, $R$, and $\teff$ refer to the parameters of a star, not of the disc. Initial stellar mass $M$ has to be changed according to alpha to ensure that it approaches 1 $\msun$ at the end of the simulation.}
  \begin{tabular}{ l l l }
  Parameter & Value & UNIT \\
  \hline \hline \noalign{\smallskip}
  $\alpha$ & $10^{-4}$ - $10^{-2}$ & $-$ \\
  $v_\mathrm{frag}$ & $10^{2}$ - $10^{3}$ & \SI{}{\centi\meter\per\second}\\
  $d/g$ & 0.01965 &  \\
  $M$ & depending on alpha & \\
  $M_\mathrm{disc}$ & 0.1$M$ & $\msun$ \\
  $\teff$ & stellar models & K \\
  $R$ & stellar models & R$_{\odot}$ \\
  $r_{\rm min}$ & 0.05 (default) & AU \\
  $r_{\rm max}$ & 3000 (default) & AU \\
  $r_\mathrm{c}$ & 200 & AU \\
  $\rho_\mathrm{s}$ & 1.156 (default) & \SI{}{\gram\per\cubic\centi\meter}\\
  $a_0$& $10^{-5}$ & \SI{}{\centi\meter} \\
  \noalign{\smallskip}
  \hline \noalign{\smallskip}
  \label{tab:1}
\end{tabular}
\end{table}

The evolution of the gas disc is modelled using the standard $\alpha$-viscosity prescription \citep{Shakura1973}, in which the transport of angular momentum is driven by turbulent viscosity. The dimensionless free parameter $\alpha$ describes homogeneously and isotropically distributed turbulence on scales much smaller than the characteristic radius of the disc, $r_{c}$\footnote{The characteristic radius defines the effective size of the disc by setting an exponential cut-off in the disc's gas and dust density}. Observations of PP discs \citep[e.g.][]{Flaherty2016, Dullemond2018} show that the plausible range of $\alpha$ values is closing in around $10^{-4}$ to  $10^{-2}$. 

The fragmentation threshold velocity $v_\mathrm{frag}$ is the relative velocity that defines whether the dust grains will grow by sticking together or fragment. Collision leads to fragmentation of grains, if they move with velocities greater than $v_\mathrm{frag}$. Our choice of the fragmentation velocity is motivated by laboratory experiments of dust and water ice grains \cite[e.g.][]{Gundlach2015}, however these show that the exact fragmentation velocity depends on the shape and composition of the dust grains and is still under debate \citep{Musiolik2019, Steinpilz2019}. The influence of both parameters, $v_\mathrm{frag}$ and $\alpha$, on the chemical evolution of PP discs is discussed in Sect. \ref{sec:acchis}. 

The temperature structure of the disc is estimated by taking passive irradiation and viscous heating into account. TPP also includes radial drift of dust grains that implies faster accretion of solids onto the star and, hence, the decrease of dust mass in the disc with time. This is supported by observational studies of disc masses in different evolutionary stages, which favour strong trends where younger discs possess significantly larger dust mass compared to older, more evolved, discs \citep[e.g.][]{Ansdell2017,Williams2019-lifetimes}. Pressure bumps in the disc may act as traps for inward moving dust particles \citep{Whipple1972}, thereby slowing down the drift of grains and prolonging the lifetime of dust. However, only a small fraction of discs ($5-10\%$ of all discs) contain large dust masses after 1 Myr \citep{Williams2019-lifetimes}, implying that a large fraction of discs must have lost their dust component onto the central star. Therefore, we do not include pressure bumps in the model.

TPP requires a number of input parameters to be defined. These are mass ($M$), effective temperature ($\teff$) and radius ($R$) of the star, characteristic radius ($r_\mathrm{c}$), initial mass ($M_\mathrm{disc}$) and dust-to-gas ratio ($d/g$) of the disc, density of the dust particles ($\rho_\mathrm{s}$), size of the smallest dust grains ($a_{0}$), fragmentation velocity ($v_\mathrm{frag}$) and turbulence parameter ($\alpha$). The geometric size of the disc is given by the minimum $r_{\rm min}$ and maximum $r_{\rm max}$ radii, respectively. Our adopted values of these parameters are given in Table \ref{tab:1}. 

The dust-to-gas ratio corresponds to the ratio of the Sun \citep{Grevesse1993}. The disc mass of 0.1 $M$ corresponds to a typical gas mass of young protoplanetary discs \citep{Andrews2010}. The critical radius of the disc of $200$ AU corresponds to a large disc, as they are seen in the DSHARP \citep{Andrews2018} survey. The choices of stellar parameters - $\teff$, $M$, and $R$ - are described in Sect. \ref{sec:garstec}.
\subsection{Mass and chemistry of the accreted material}\label{sec:chem}
TPP does not conserve mass. The main mechanism for mass loss from the disc is the accretion of material onto the central mass, and only a minuscule fraction of less than 0.1 $\%$ is lost through the outermost boundary of the disc. This allows us to compute the properties of accreted matter as follows.

TPP supplies us with surface densities $\Sigma_{g}$ for gas and $\Sigma_{d}$ for dust as a function of the disc radius and time. We compute the mass of each phase, $M_{d}$ is mass of the dust and $M_{g}$ the mass of the gas, at a given point in time by integrating over the area of the disc:
\begin{equation}
    M_{g,d}(t) = 2\pi \int_{r_\mathrm{min}}^{r_\mathrm{max}} r \Sigma_{g,d}(t, r) dr
    \label{eq:1}
\end{equation}  
where $r_\mathrm{min}$ and $r_\mathrm{max}$ are defined in Table \ref{tab:1}. The evolution of $M_\mathrm{g,d}$ is driven by the accretion of material onto the star, therefore we simply assume that the differences between the time steps $i$ and $i+1$ represent the total accreted mass at $i+1$. Note, that the initial $M_\mathrm{g,d}$ at time zero do not depend on $r_\mathrm{min}$ and $r_\mathrm{max}$, but have been set by $M_{\rm disk}$ and $d/g$.

We assume that the metallicity of the dust phase is exactly one and the metallicity of the gas phase is exactly zero. Hence, we can calculate the metallicity of the material accreted between the timesteps as follows:
\begin{equation}
 Z_\mathrm{accr} = \frac{\dot{M_{d}}}{\dot{M_{d}} + \dot{M_{g}}}
 \label{eq:2}
\end{equation} 
We think that this approach is appropriate in the context of our work, as meteorites in the solar system are known to contain negligible fractions of H and He \citep{Lodders2003}.

Setting more precise constraints on the evolution of the disc chemistry requires a detailed chemical network including radial drift and growth of these grains, as well as evaporation and condensation of these grains at ice lines \citep[e.g.][]{Booth2017}. However, providing a detailed model of the exact composition of the material is beyond the scope of this work.
\subsection{Stellar evolution models}\label{sec:garstec}
Stellar models have been computed with the \texttt{Garstec} code \citep{Weiss2008}. We calibrated the mixing length parameter, initial metallicity and helium abundance on the Sun, using the solar metal abundance given by \citet{Grevesse1998} (A(Fe) $= 7.50$, (Z/X)$_{\rm phot} = 0.0229$ dex). From the calibration we obtain $\alpha_{\rm MLT}= 1.811$ and the initial solar helium abundance Y$_{\rm ini}= 0.26896$ and metallicity Z$_{\rm ini}= 0.01876$. Due to the effects of microscopic diffusion, the initial solar composition corresponds to [Fe/H] $=$ +0.06.

For the present work, we have computed a dense grid of stellar models in the mass range $0.6 \msun \leq M \leq 6 \msun$ in steps of 0.02 up to 3 solar masses and in steps of 0.04 from 3 to 6 solar masses. The initial composition of the models is computed assuming a cosmic helium-to-metal enrichment ratio of $\Delta Y/ \Delta Z = 1.1$ anchored to the initial solar composition and a cosmological $Y_{\rm SBBN}=0.2487$ \citep{Steigman2007}.

\begin{figure}
  \centering
  \includegraphics[scale=0.9]{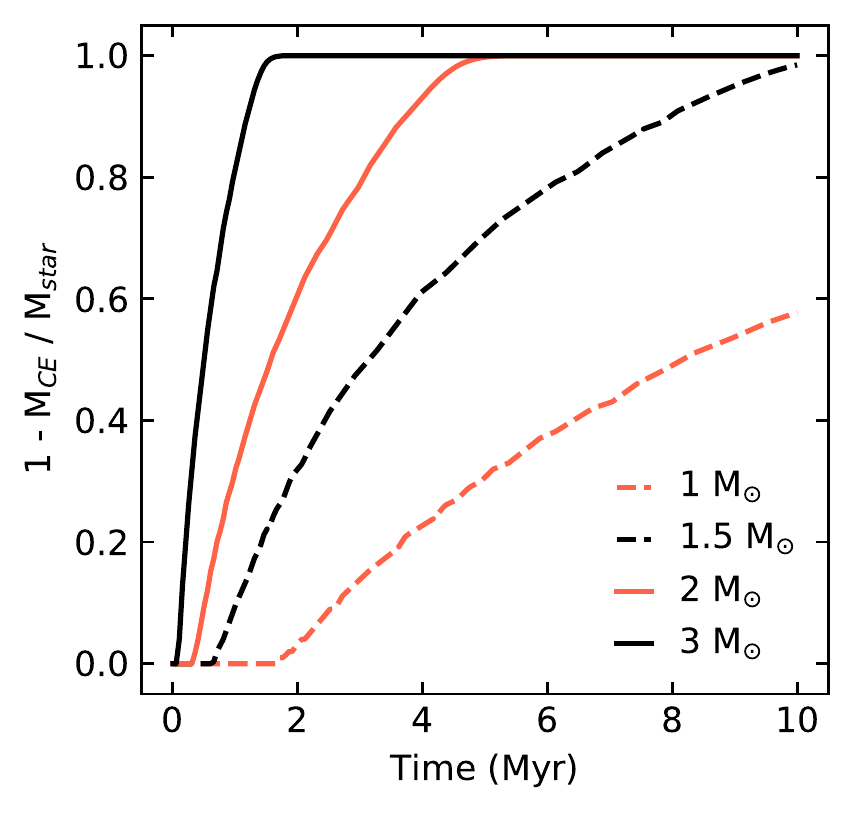}
  \caption{Evolution of the convective envelope extension in pre-MS stars of 1 to 3 solar masses as a function of the stellar age.}
  \label{fig:envelopes}
\end{figure}

Figure \ref{fig:envelopes} shows the evolution of the mass inside the convective envelope $M_\mathrm{CE}$ for stars of different initial mass ($M_{\star}$) during the early stages of evolution. The convective envelope recedes faster for more massive stars. $M_\mathrm{CE}$ on the main sequence depends primarily on the $\teff$ of the star, that is on the stellar mass and metallicity. For solar metallicity, which is indicative of the models used in this work, stars above 1.5~$\msun$ lose their convective envelopes completely (except for a very thin envelope $<10^{-5}\msun$), whereas convective envelopes for a solar mass never recede below 0.02~$\msun$ during the star's life on the MS.

\subsection{Accretion histories}\label{sec:acchis}
The two key parameters that impact the evolution of PP discs are the turbulent viscosity parameter $\alpha$ \citep{Shakura1973} and the fragmentation velocity $v_\mathrm{ frag}$ that sets the lower limit for the growth of dust. Since it is not possible to set tighter constraints on these parameters at this stage, we explore a range of values for both $\alpha$ and $v_\mathrm{frag}$ covering the parameter space reported in the literature. In particular, we vary $v_\mathrm{frag}$ from 10$^2$ to 10$^3$ cm s$^{-1}$ and $\alpha$ from $10^{-4}$ to $10^{-2}$  (Sect. \ref{sec:tpp}). 

\begin{figure}
  \centering
  \includegraphics[scale=0.9]{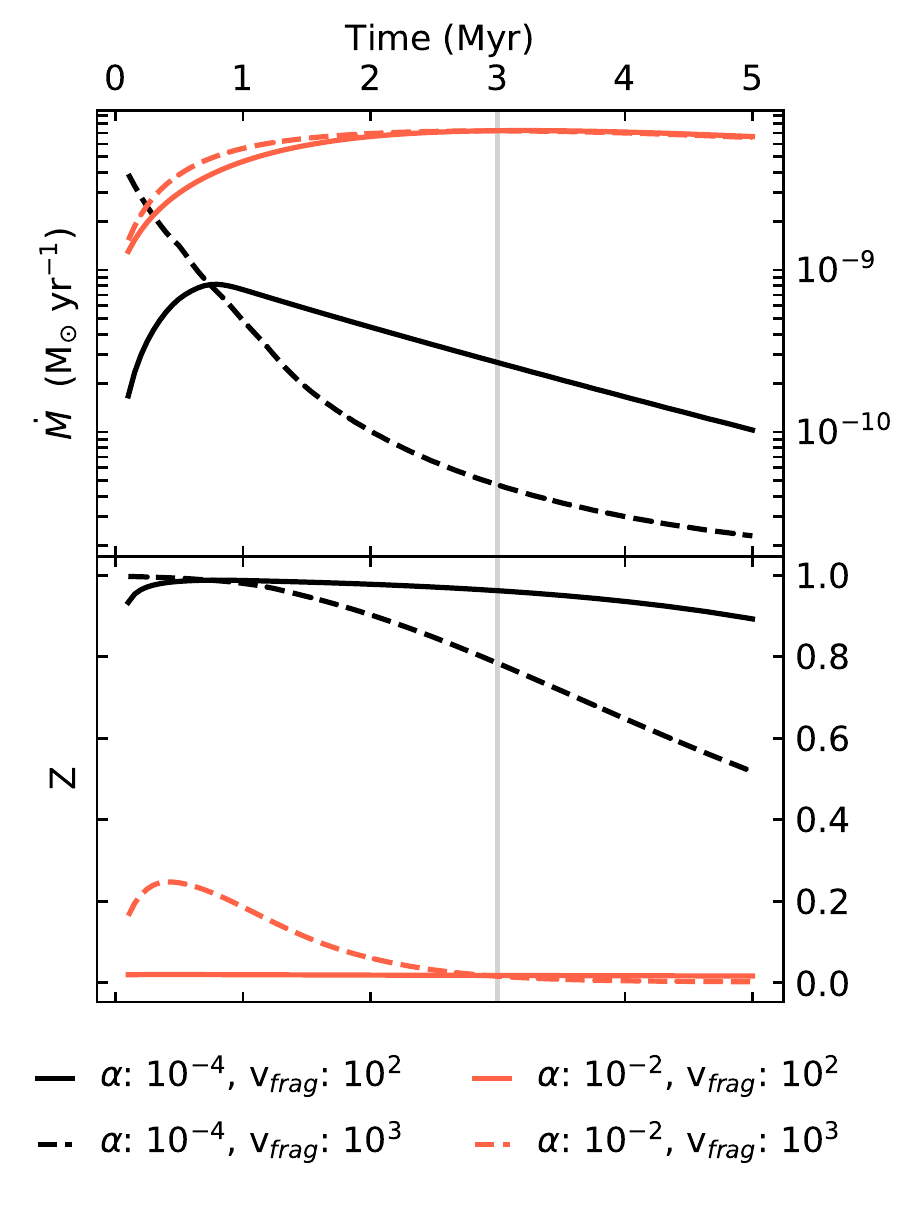}
  \caption{Accretion histories depending on input parameters $\alpha$ and $v_\mathrm{frag}$. The upper plot shows the total accretion rate (gas+dust) onto the central star as a function of time. The bottom plot shows the metallicity Z of the accreted material as a function of time. All shown accretion histories result in a solar mass star after 3 Myr indicated by the vertical grey line. }
  \label{fig:accretion}
\end{figure}

The temporal evolution of the mass accretion rate and of the metallicity in the accreted matter for four different combinations of the two parameters are shown in Fig. \ref{fig:accretion}. Clearly, even within the limited parameter space explored in this example, we find a broad variety of accretion scenarios. The mass accretion rates range from $10^{-10}$ to $10^{-8}$ $\Msun{}$yr$^{-1}$ and show very different behaviours, depending on the values of $\alpha$ and $v_{\rm frag}$. A high $\alpha$ parameter, meaning a very turbulent disc, leads to rapid gas accretion and, therefore, to a low metallicity of the accreted material. In the opposite case, low $\alpha$ slows gas accretion down, which allows the dust to dominate the early accretion. This effect is caused by the changing grain growth and drift properties with different viscosities. At low viscosity, grains grow larger and thus drift faster inwards, leading to a dust dominated accretion history, while at a high viscosity, grains stay small (due to the higher turbulent velocities limiting their growth) and are thus drifting slowly, resulting in a gas dominated accretion history. Thus, the metallicity $Z$ of the accreted material covers the full range from zero to one. That is, the exact amount of metals that are accreted strongly depends on the choices of the disc and dust growth model.
We use these results to develop two extreme cases for the chemical evolution of accreted material and to investigate how accretion can influence the metal abundances in stellar atmospheres (Sect. 2.5).
\subsection{Chemical evolution of stellar convective envelopes}\label{sec:chemconv}
The surface metallicity of a star can be altered by accretion in several different ways. The two most obvious extreme scenarios are the metal-rich (dust only) and  metal-poor (gas only) accretion. These scenarios, in practice, encapsulate the entire diversity of accretion scenarios predicted by the variation of $\alpha$ and  $v_{\rm frag}$ parameters, that control the general evolution of PP discs (Sect. 2.4). It is therefore reasonable to adopt two limiting cases of accretion: accretion of dust (Z $=$ 1) and accretion of the gas (Z $=$ 0).

If the accreted material is metal-rich, compared to the composition of the convective envelope, the surface metallicity will be increased. If the accreted material is metal-poor, metals in the convective envelope will be diluted, leading to a decrease in metallicity. The efficiency of the dilution process depends on the extension of the convective envelope, which (in the selected mass range) recedes within the first $\sim 10$Myr. This can lead to an appreciable variation of the chemical composition of stellar photospheres.

We assume instantaneous accretion and mixing throughout the convective envelope of a star at a fixed time $t$. Since the convective envelope is monotonically receding, this gives the limiting change in metallicity for all accretion histories within a timescale $t$. We also neglect perturbations of the mechanical equilibrium structure of the star that could possibly be caused by the transfer of energy and momentum from the disc. 

In each of the two afore-mentioned scenarios, the final metallicity of the convective envelope is given by:

\begin{equation}
    Z_\mathrm{CE, max} = \frac{M_\mathrm{CE}(t) \, Z_\mathrm{ini} + M_\mathrm{disc} \, Z_\mathrm{ini}}{M_\mathrm{CE}(t) + M_\mathrm{disc} \, Z_\mathrm{ini}}
    \label{eq:Zmax}
\end{equation}

\begin{equation}
    Z_\mathrm{CE, min} = \frac{M_\mathrm{CE}(t) \, Z_\mathrm{ini}}{M_\mathrm{CE}(t) + M_\mathrm{disc} \, (1-Z_\mathrm{ini})}
    \label{eq:Zmin}
\end{equation}

\noindent where $M_{C\!E}$ is the mass of the stellar convective envelope, $M_{disc}$ is the initial mass of the disc before accretion, $t$ the time at which accretion occurs and $Z_{ini}$ is the initial metallicity, which is assumed to be the same for the star and its  disc. Eq. \ref{eq:Zmax} describes metal-rich accretion and Eq. \ref{eq:Zmin} describes metal-poor accretion. 

One has to keep in mind that in each of the two scenarios the mass of the star itself changes. This has to be accounted for when comparing the results with observations from clusters as we will do in the next section.
\section{Results}\label{sec:res}
We begin the discussion with comparing our predictions for the Sun with its present-day photospheric chemical composition (Sect. \ref{sec:sun}). We then provide quantitative predictions for the accretion-induced changes in surface properties of stars across a range of ages and metallicities. We compare our model predictions with the observed colour-magnitude diagrams (CMD) of open clusters in the Gaia DR2 (Sect. \ref{sec:clust}) and discuss whether accretion from the PP discs is a viable mechanism to explain the broad MS turn-offs in clusters.

\begin{figure}
  \centering
  \includegraphics[scale=0.9]{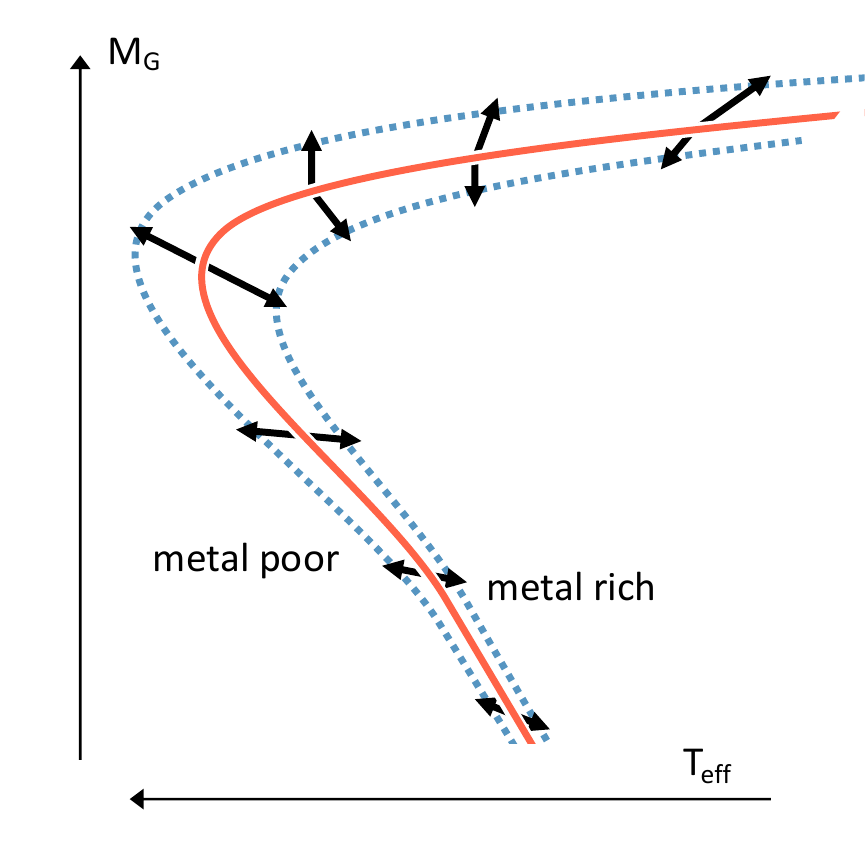}
  \caption{Qualitative metallicity variation effects on the isochrone in the absolute magnitude ($M_{\rm G}$) vs $T_{\rm eff}$ plane.}
  \label{fig:Cartoon}
\end{figure}
\subsection{The peculiar composition of the Sun}\label{sec:sun}
It is known that the chemical composition of the Sun is somewhat peculiar. The Sun appears to be more metal-poor in refractory elements compared to the stars in the solar neighbourhood \citep{Ramirez2014,Adibekyan2014,Nissen2016,Bedell2018}.
Recently, the composition of the solar photosphere was revised using 3D non-local thermodynamic equilibrium (NLTE) models by \citep{asplund2005,asplund2009} and \citet{Caffau2011}.  The new metallicity, and in particular the oxygen abundance, is significantly lower than that given in the old compilation from \citet{Grevesse1998}, who relied on simpler 1D LTE models. The standard solar models that employ the 3D NLTE abundances yield the present-day structure of the Sun that is significantly different from inferences based on direct helioseismology measurements \citep[e.g.][]{Bahcall2005, Castro2007, Serenelli2009, Buldgen2017}. Various theories have been put forward to explain the mismatch, including energy transport by dark matter particles in the Sun or resonant emission of transversely polarised hidden photons (see e.g. \citealt{Vincent2015,Sokolov2020}). We explore whether our accretion scenario could potentially help to resolve the problem.

The Sun was fully convective for $\approx 2$ Myr during the first stages of its life. The lifetime of the solar nebula is constrained to $3-4$ Myr by meteorites \citep{Connelly2012,Wang2017}. We model accretion from the proto-solar disc onto the early Sun assuming that accretion happened within $3$ Myr and the protoplanetary disc mass was $10\%$ of the solar mass. Applying the two limiting case scenarios, as described in Sect. 2.5, we find that solar photospheric metallicity $Z$ could have indeed changed substantially - by $\sim 10 \%$ under the influence of accretion. In the case of metal-rich accretion (Eq. \ref{eq:Zmax}), the present-day metallicity of the solar photosphere is $Z_\mathrm{max} = 0.02004$, which is $11.33\%$ above the initial value. The initial solar mass should have been roughly 0.998 $\Msun$. In the case of metal-poor accretion (Eq. \ref{eq:Zmin}), we obtain a lower limit on the surface metallicity of $Z_\mathrm{min} = 0.01627$ which is $9.61\%$ below the initial solar metallicity. The initial solar mass should have been roughly 0.911 $\Msun$.

Thus, the change in the photospheric metallicity of the Sun is significant enough to impact the discussion of the solar modelling problem \citep{Castro2007, Guzik2010, Serenelli2011,Buldgen2019}. The estimate by \citet{asplund2009}, Z$=0.0134$, cannot be explained by our models. However, the solar metallicity reported by \citet{Caffau2011}, Z$=0.0153$, is, within the uncertainties of the data, consistent with our metal-poor accretion model. This implies that our scenario could, in principle, explain the mismatch between the present-day composition of the Sun and the SSMs constrained using helioseismology, alleviating the need for other exotic explanations. Detailed solar model calculations accounting for the specific accretion histories can be used in order to confirm this.
\subsection{Influence on observable properties of stellar populations}\label{sec:clust}
As discussed earlier, accretion of matter from the PP disc onto the star changes its total mass and metallicity. For low-mass stars, the material is (almost instantly, compared to evolutionary timescales) diluted across the entire convective envelope, hence the metallicity effect will strongly depend on the size of the CE. The higher the mass of a star, the more pronounced are the changes with respect to the initial abundances. This constant addition of matter and re-adjustment of the chemical composition in the CE, furthermore, leads to changes in the physical structure of a star and its surface  properties, and, consequently, in the position of a star in the plane of observable quantities, $\teff$ and $\logg$.

In what follows, we will explore the consequences of stellar evolution modulated by accretion from PP discs for the position of stars in colour-magnitude diagrams. We will quantify the influence of accretion on evolutionary tracks of stars with different masses and initial metallicities and assess the modulation of surface metallicity against age. We will also compare the predictions of our models with several Galactic open clusters, for which precision astrometry from Gaia DR2 is now available, in order to test the viability of our scenario.
\begin{table}[ht!]
\centering
  \caption{Nearby open clusters used in this work. Data adopted from \citet{Babusiaux2018}.}
  \begin{tabular}{ l r r r }
  \hline \hline \noalign{\smallskip}
  Cluster & Parallax & Distance & Metallicity (Z) \\
   & mas & pc &  \\
  \hline \hline \noalign{\smallskip}
  Hyades & 21.052 & 47.5 & 0.020\\
  Coma Ber & 11.640 & 85.9 & 0.017\\
  Pleiades & 7.364 & 135.8 & 0.017\\
  Praesepe & 5.371 & 186.2 & 0.020\\
  $\alpha$ Per & 5.718 & 174.9 & 0.020\\
  Blanco 1 & 4.216 & 237.2 & 0.017\\
  \noalign{\smallskip}
  \hline \noalign{\smallskip}
  \label{tab:2}
\end{tabular}
\end{table}
%
%----------------------
\subsubsection{Observations of open clusters}
In this work, we select all nearby open clusters with the age of $\lesssim 0.8$ Gyr released with the Gaia DR2 \citep{Babusiaux2018}. In addition, we exclude some of the clusters that contain too few members. This leaves us with six open clusters with ages from $\sim 70$ Myr ($\alpha$ Per) to $\sim 800$ Myr (Hyades) at heliocentric distances of $\sim 50$ to $\sim 250$ kpc \citep{Babusiaux2018}. Their basic parameters are listed in Table~\ref{tab:2}. The cluster membership was determined via an iterative procedure from the analysis of the positions of stars, their proper motions, and parallaxes resulting in a very reliable classification \citep[][their Appendix A.1 and Table A.2]{Babusiaux2018}. The proximity of clusters is essential, as, owing to high accuracy of parallaxes and proper motions, this minimises the problem of fore- or back-ground contamination. We illustrate the uncertainties of Gaia measurements for two clusters in our sample, Praesepe and Blanco 1, in Figs. \ref{fig:ErrBlanco} and~\ref{fig:ErrPraesepe} in Appendix. 
\begin{figure*}
  \centering
  \hbox{
  \includegraphics[scale=0.65]{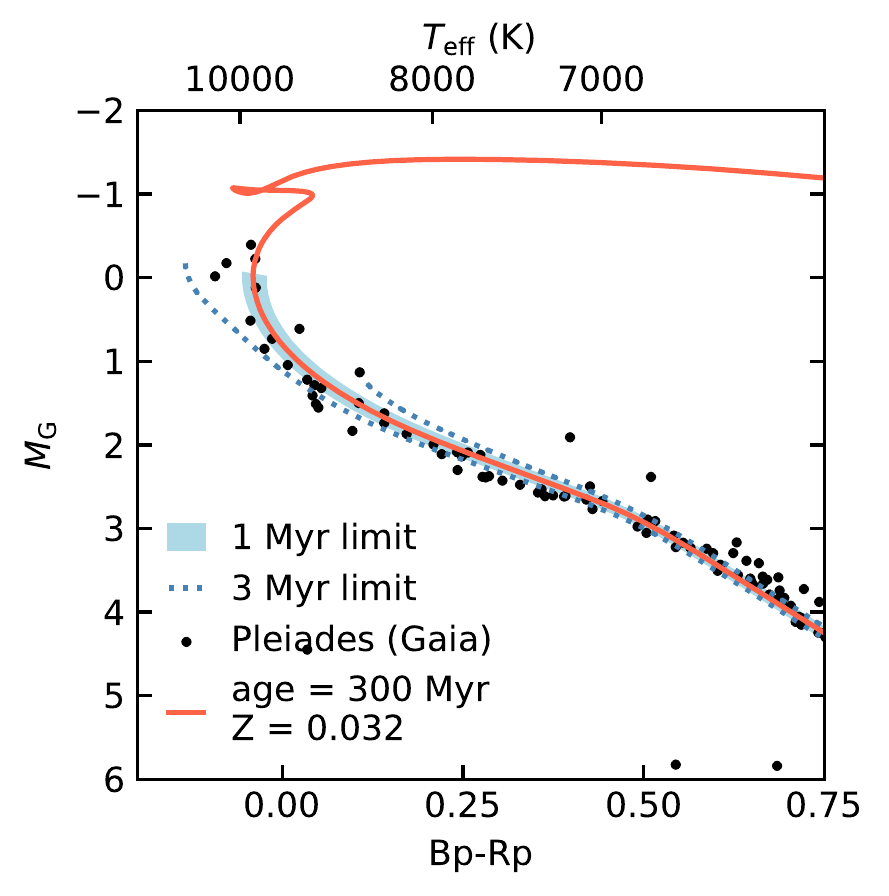}
  \includegraphics[scale=0.65]{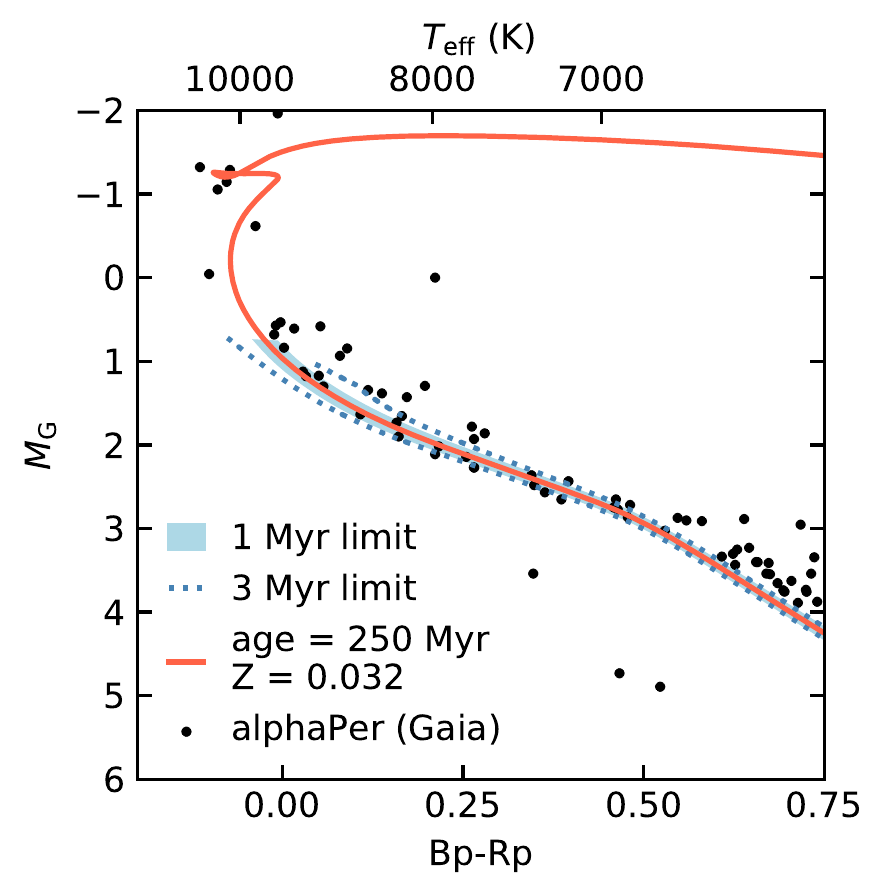}
  \includegraphics[scale=0.65]{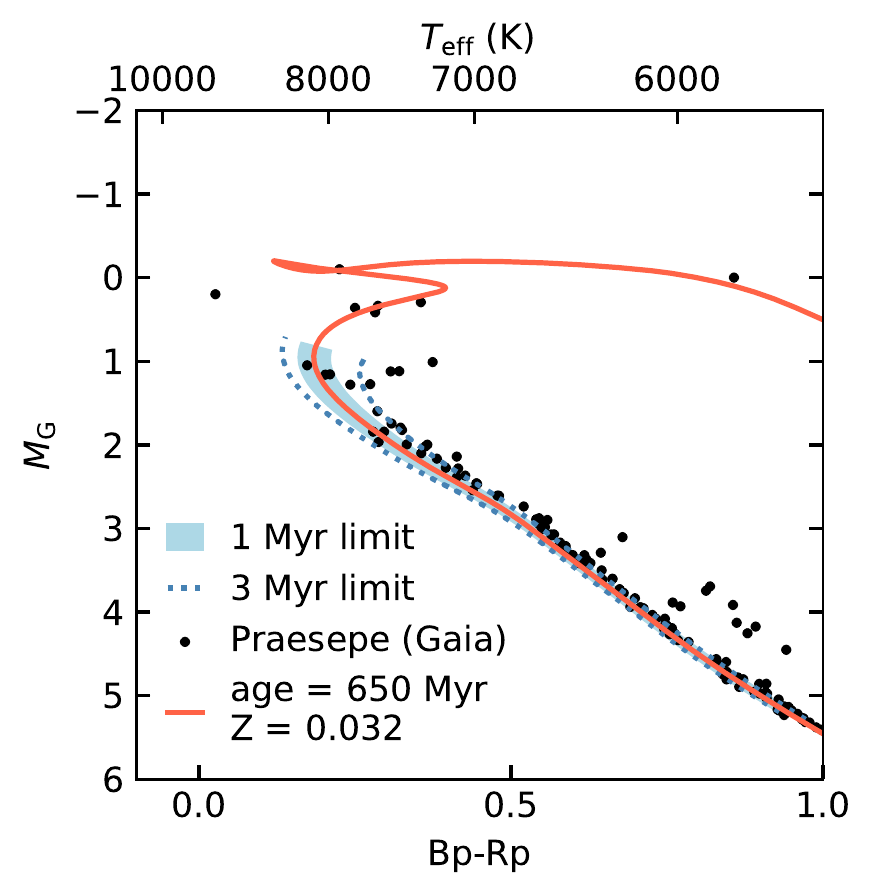}
   }
  \hbox{
  \includegraphics[scale=0.65]{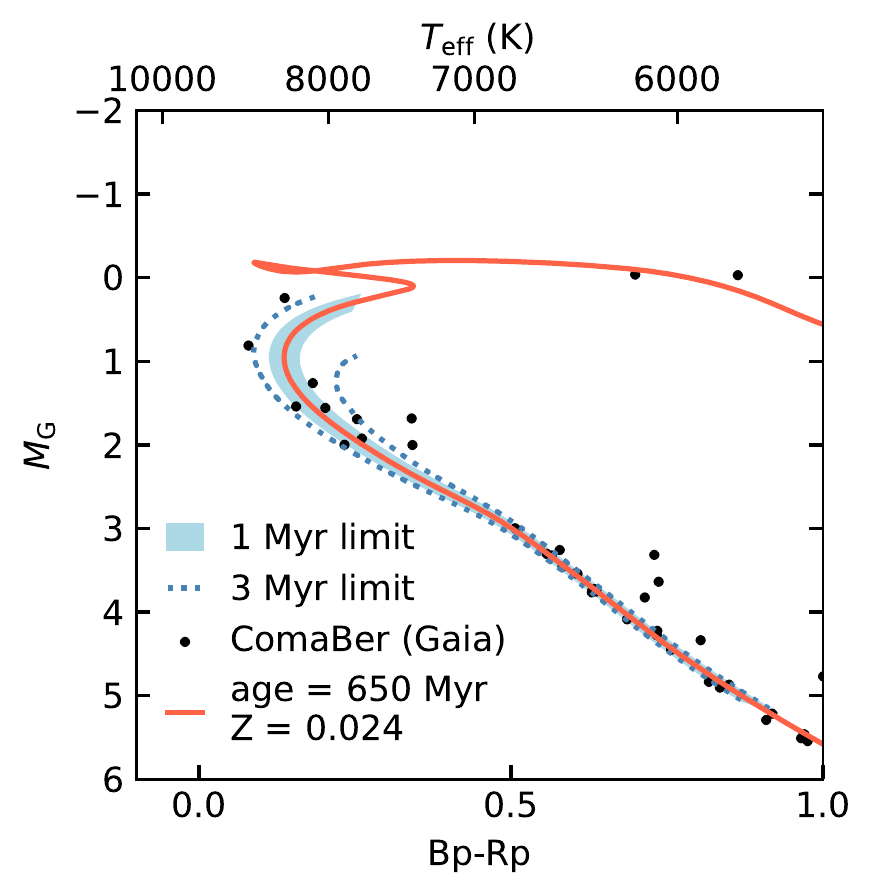}
  \includegraphics[scale=0.65]{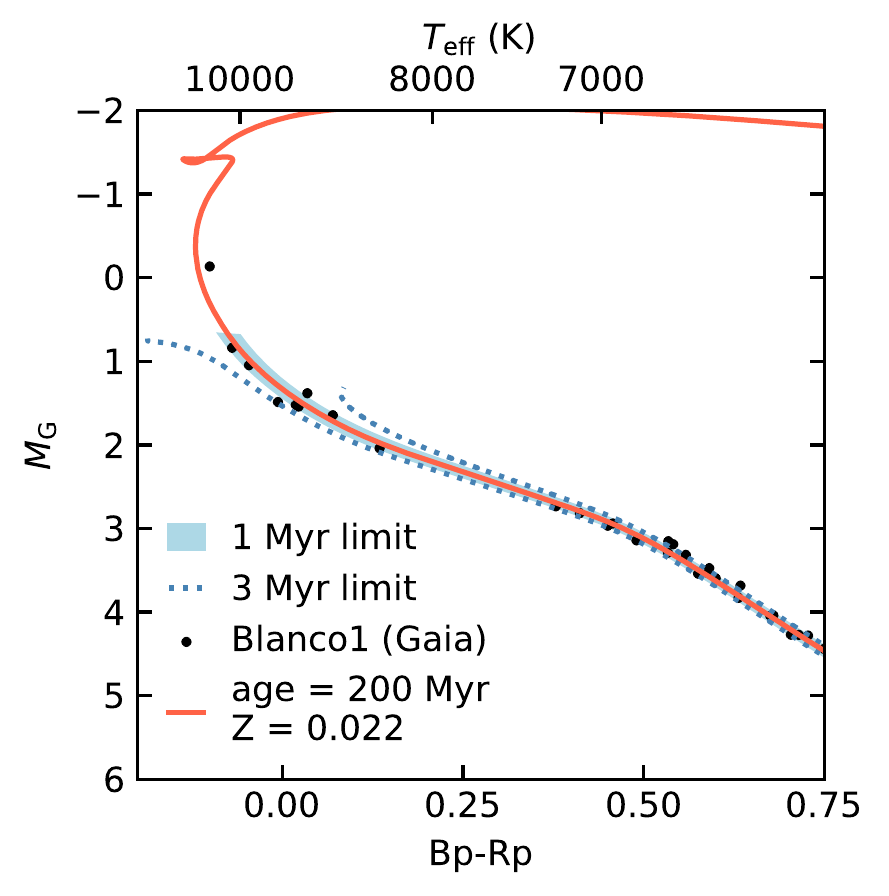}
  \includegraphics[scale=0.65]{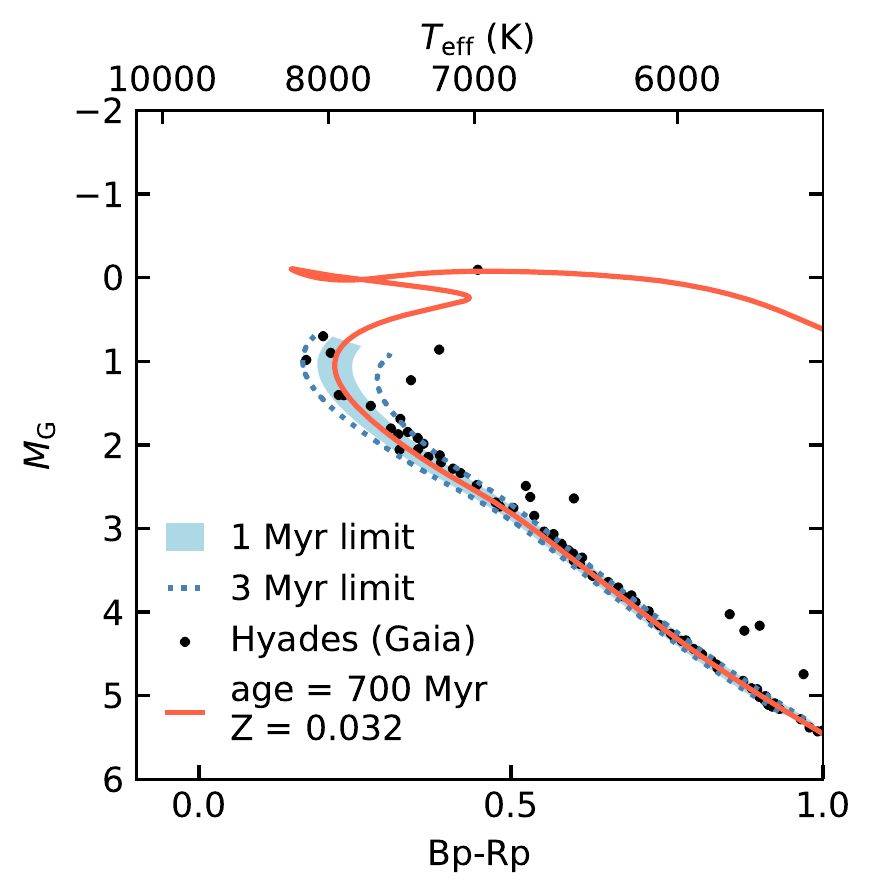}
  }
  \caption{Hertzsprung-Russell diagrams of nearby open clusters from Gaia DR2. The blue shaded area depicts the range of Bp-Rp that could be explained by different accretion histories for a disc lifetime of 1 Myr. The area between the two dotted blue lines shows the same effect, but for a disc lifetime of 3 Myr.}
  \label{fig:HR1}
\end{figure*}

The clusters are shown in Fig.~\ref{fig:HR1}. Symbols represent the data from Gaia DR2. Representative isochrones are over-plotted. These isochrones are not to be viewed as best-fit isochrones, as we made no attempt to fit them statistically, but chose to show a representative model that describes the cluster main-sequence. This procedure does not impact our results and conclusions, because we are only interested in the dispersion of stars at the cluster MSTO and this parameter does not depend on the choice of the model, as long as the cluster's main-sequence is reproduced. 
\subsubsection{Impact of protoplanetary disc accretion on cluster morphology}
It is interesting that despite an overall consistency between the observed positions of stars in the clusters and the theoretical isochrones, there is a significant number of outliers in the observed data and their fraction increases at the MSTO. This peculiarity has already been reported in the literature \citep{Marino2018, Cordoni2018} and is termed extended main-sequence turn-off (hereafter, eMSTO). 
A number of physical scenarios have been put forward to explains eMSTOs, such as rotation \citep{Marino2018b, Sun2019} and multiplicity \citep[e.g.][]{Bastian2009,DAntona2016,Cordoni2018}. Here we explore whether this peculiar effect can be potentially explained with our scenario of accretion from protoplanetary discs.

Using equations \ref{eq:Zmax} and \ref{eq:Zmin} we can compute minimum and maximum metallicities as a function of $M_{\star}$ and $M_\mathrm{disc}$ assuming some $Z_{\rm ini}$. Linking metallicity to stellar mass we computed two extreme isochrones for each cluster of a given age and metallicity. To probe the effect of the disc lifetime, we furthermore explored two cases. One case represents the accretion up until the age of $1\,$Myr, the other corresponds to extended accretion up to $3\,$Myr, which is equivalent to a longer-lived PP disc.

The results of our model for all our clusters are shown in Fig. \ref{fig:HR1}\footnote{Effective temperatures on top axis of Fig. \ref{fig:HR1} are taken from:  \url{http://www.pas.rochester.edu/~emamajek/EEM_dwarf_UBVIJHK_colors_Teff.txt}.}, where the orange line corresponds to the fiducial cluster isochrone, whereas the shaded area and the dotted lines represent the range of possible displacements of stars within this cluster given the assumed accretion history, that is, short- and long-lived PP discs. We overplot the observed clusters with very accurate parallaxes and photometry from Gaia DR2. Accretion of matter from the PP disc leads to changes in the evolutionary track of a star and, consequently, in the position of a star in  the Bp-Rp - $M_G$ plane\footnote{Note that we prefer to work in the Bp-Rp - $M_G$ plane, as colours and magnitudes are extremely well measured by Gaia, whereas the GDR2 $\teff$ estimates are very uncertain. Also, owing to the proximity of clusters, reddening is not a problem for our study.}. Stars that experienced the accretion of metal-poor material (gas) from their discs are shifted to the left, towards bluer colours (or higher $\teff$), whereas the objects that underwent the accretion of metal-rich material are displaced to the right, towards redder colours (or lower $\teff$).  The displacement is not the same along the isochrone because it depends on the extension of the convective envelope and, hence, on the stellar mass.

The net result is that the cumulative distribution of stars in a cluster \textit{can no longer be described by a single isochrone}. Instead, stars show a significant dispersion around the characteristic locus that corresponds to a best-fit model. This scatter gradually increases and attains maximum at the cluster MSTO point, because MSTO stars have the smallest convective envelopes. Extended accretion, as, for example, caused by longer-lived discs, increases the intra-cluster MSTO scatter. Consequently, the loci corresponding to 3 Myr discs are significantly wider compared to the loci computed for shorter-lived 1 Myr discs. 

Comparing our results with the observations of nearby open clusters, we find that the MSTO scatter predicted by our model indeed matches the data. Every cluster studied in this work, except Blanco 1, shows the characteristic observed dispersion of $\sim$ 100 to 200 K at the MSTO (corresponding to $0.05$ mag in Bp-Rp), which is qualitatively consistent with the simulated loci (Fig. \ref{fig:HR1}). The theoretical MSTO spread is large for younger clusters, but it is not as significant for older systems. The observed dataset for Blanco 1 is too sparse and it does not allow to quantify its MSTO dispersion. At this stage, it is not possible to distinguish between long- or short-lived discs, because the samples of cluster members are still very small and selection functions are difficult to quantify. Altogether, this may lead to a biased measure of the observed MSTO width.
Nonetheless, our results provide enough evidence that accretion induced dispersion in $\teff$ and metallicity of stars could match the width of eMSTOs found in nearby open clusters.

\subsubsection{Evolution of the intra-cluster metallicity spread}

\begin{figure}[ht!]
  \centering
  \includegraphics[scale=0.8]{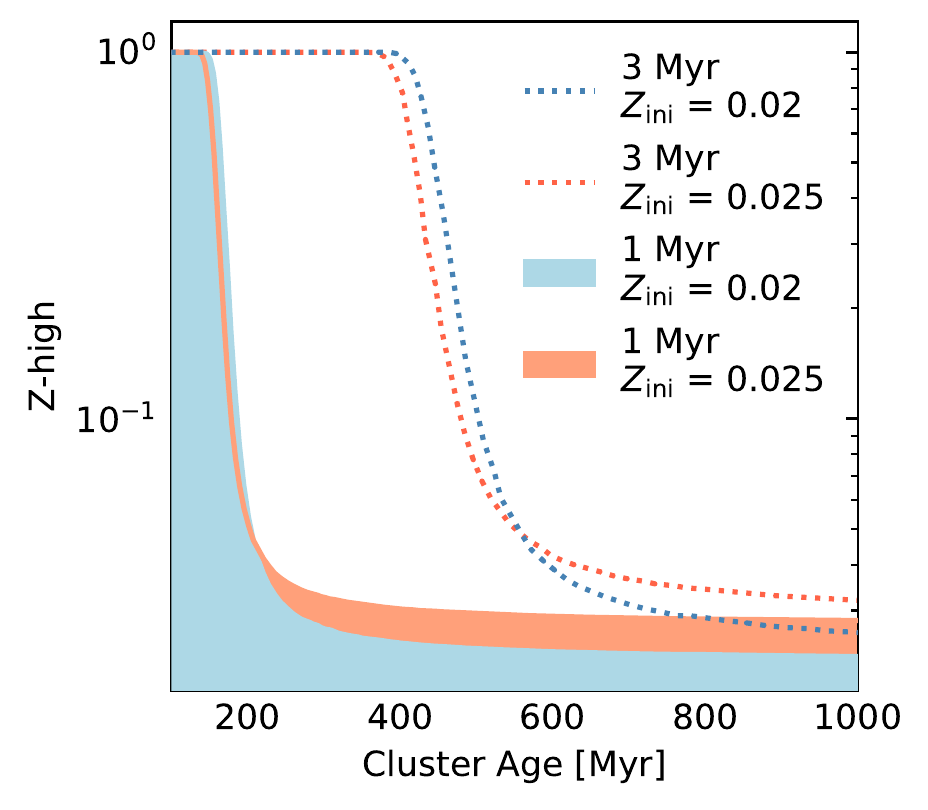}
  \includegraphics[scale=0.8]{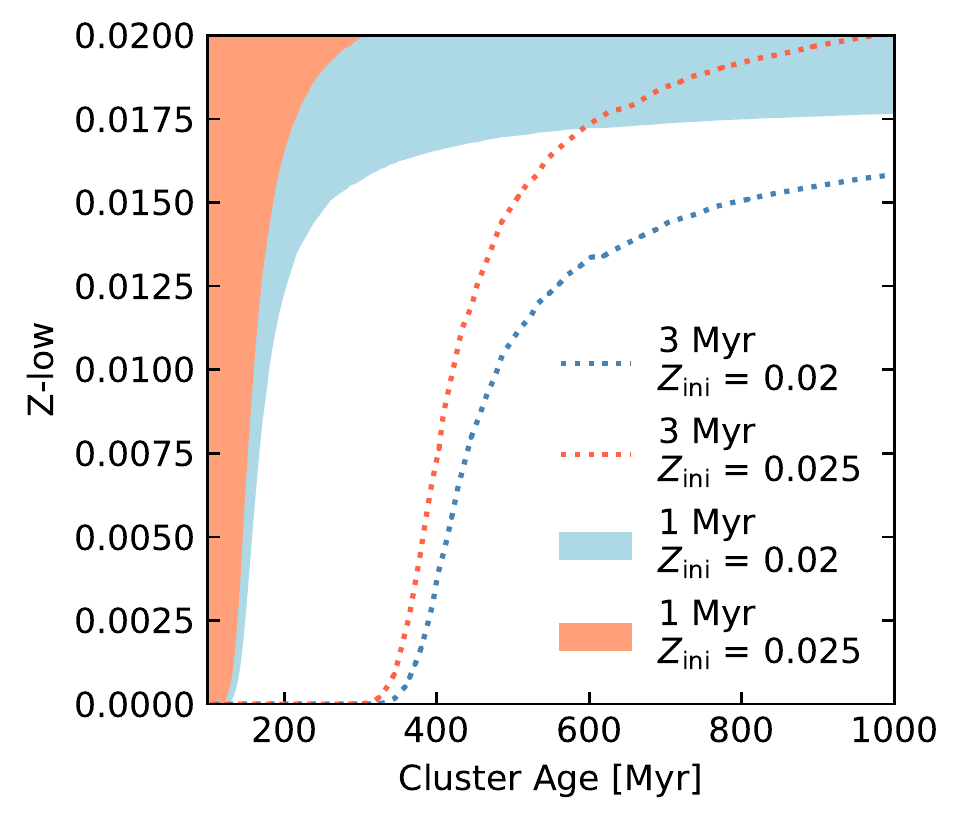}
  \caption{Expected high-metallicity (top panel) and low- metallicity (bottom panel) limits for stars  at the main-sequence turn-off as a function of the age of the cluster. The shaded areas mark everything within the limits for an accretion history completed within 1 Myr, whereas the dotted lines extend the limits for accretion within 3 Myr. }
  \label{fig:zacc}
\end{figure}

As mentioned above, our scenario also predicts a systematic variation of metallicity across the MSTO of a cluster. Figure  \ref{fig:zacc} illustrates the metallicity effect against the cluster age. Clearly, the MSTO stars in very young clusters, age $< 200$ Myr ($M_{\rm TO} > 3 \Msun$), are most affected by accretion, as illustrated by shaded areas in these figures. These stars fully lack convective envelopes, and hence may inherit either a completely metal-dominated atmosphere (Fig. \ref{fig:zacc}, top panel) or a pure hydrogen-helium atmosphere (Fig. \ref{fig:zacc}, bottom panel). 

Generally, the longer the lifetime of the disc, the more prominent is the effect. For example, in case of accretion lasting 3 Myr, also older clusters, with ages of up to $500$ Myr ($M_{\rm TO} \sim 2.5 \Msun$), start showing this effect. Also the initial metallicity of the stars in the cluster, $Z_{\rm ini}$, plays a role, especially in the case of metal-poor accretion. So, for example, for $Z_{\rm ini}$ of 0.02, the stellar atmosphere quickly becomes rich in H and He, compared to its original chemical composition and this depletion of metals, of the order $\sim 0.005$, is seen for all clusters regardless of their age. In contrast, in the case of $Z_{\rm ini}$ of 0.025, only the most massive stars $> 3.3 \Msun$, which can only be found in very young clusters, show a signature of metal depletion in their atmospheres. This is due to the fact that the convective envelopes of metal-poor stars recede more rapidly than the envelopes of stars of higher initial metallicity. Therefore their envelopes are smaller at the time of accretion and the change in their their surface metallicity is more prominent.

The metallicity gradient across the cluster MSTO is a very strong test of our scenario. However, notwithstanding major progress in observational studies of open clusters (e.g. the Gaia-ESO survey), there are still major limitations that make it impossible for us to use the observed samples to test our models. The main problem is that cluster membership is often based not only on stellar kinematics, but also on the metallicity of stars \citep[e.g.][]{Blanco-Cuaresma2015, Blanco-Cuaresma2018}; so the samples will be, by construction, devoid of objects with metallicities deviating from the average value of the cluster. Secondly, the interesting nearby clusters lack a thorough spectroscopic analysis. \citet{Liu2016} and \citet{Dorazi2020} confirm the chemical inhomogeneity of Hyades and Praesepe, but their results are based on only a few MS-stars and they cannot be used as a statistically significant sample. \citet{Takeda2013} explored chemical composition of 68 stars in Hyades.  \citet{Casamiquela2019} derived LTE abundances for 22 elements for $62$ Hyades and $22$ Praesepe stars. However, both of these studies are limited to F-G type stars and do not probe the TO of the clusters. More distant open clusters in the interesting regime of ages and metallicities have only been studied with photometry \citep{Marino2018,Cordoni2018}. \citet{Magrini2014} provide abundances for clusters Trumpler 20, NGC 4815 and NGC 6705, but their sample  only contains 21, 13 and 5 members, respectively. Finally, older clusters (such as globular cluster systems or old open clusters) are extensively studied by means of high-resolution spectroscopy \citep[e.g.][]{Gruyters2014, Gruyters2016, Gao2018, Kovalev2019}, however our model suggests that the accretion induced scatter will be very small, because of their high age and the resulting deep convective envelopes of their TO members.

\subsubsection{Detailed stellar evolution with accretion}
In our attempt of placing stars with an altered surface composition on a colour-magnitude diagram, we used stellar structure models with a uniform composition, albeit modified by accretion. We assumed that the evolution and surface parameters of a star - its colour and magnitudes - only depend on the chemical composition of its outermost layers and its age. This was done for the sake of time and resources, because self-consistent evolution of stellar structure that takes surface accretion into account is computationally very expensive. However, the validity of this assumption has to be checked.

To explore the influence of heterogeneous composition on stellar evolution, we compared evolution models of stars with a uniform composition to evolution with heterogeneous composition computed using the \texttt{LPCODE} \citep{Althaus2010b, Renedo2010, Salaris2013}. We tested how the physical parameters of a Z$ =0.02$ star react to doubling or halving the metallicity of its 3\% upper mass. As expected, the chemical composition at the surface adjusted the effective temperature of the star, but its absolute magnitude and its lifetime were not affected. The effective temperature was modified by $\sim 300$K for stars of 1.5$\Msun$ and by $\sim 500$K for 3$\Msun$ relative to the model with a uniform chemical composition. Although this change is slightly smaller compared to the predictions of our simple model, the test confirms that random changes in surface metallicity caused by accretion from PP discs might be responsible for eMSTOs in open clusters.

\section{Conclusions}\label{sec:conclusions}
In this work, we explore whether accretion of material from the protoplanetary disc can lead to significant changes in the surface physical parameters (metallicity and $\teff$) of the central star, with respect to its initial chemical abundance pattern set at birth.

We employed the Two-Pop-py code \citep{birnstiel2011dust} to model the evolution of the disc and the \texttt{Garstec} code \citep{Weiss2008} to model stellar evolution. We assumed instantaneous accretion and mixing throughout the stellar convective envelope, and neglected perturbations of the mechanical structure of a star. We find that the accretion histories derived with the Two-Pop-Py code are extremely diverse, both in terms of the mass accretion rate and the dust-to-gas ratio in the accreted matter. Even within the conservative limits on the key disc parameters, the turbulent viscosity parameter $\alpha$ and the fragmentation velocity $\nu_{\rm frac}$, the central star may undergo either a fast early accretion of (metal-rich) dust, a slow extended accretion of (metal-poor) gas, or some combination thereof. However, the accretion of material onto the star evolves on similar timescales as the evolution of stellar structure. This result demands accretion to have an influence on the structural evolution of the star and vice versa. We encapsulate the entire range of plausible accretion scenarios by assuming two limiting cases - the metal-free and the metal-rich accretion -, which in the simulations reported in this work.

First, we apply our method to the evolution of the Sun. We find that, although the accretion potentially affects the solar surface metallicity, the effect is not large. Starting from the old solar composition by \citet{Grevesse1998}, $Z = 0.018$, which we assume to be representative of the proto-Sun, we obtain $Z = 0.01627$ and $Z = 0.02004$ for the metal-poor and metal-rich accretion scenarios, respectively. The value reported by \citep{Caffau2011}, $Z = 0.0153$, is consistent with our lower estimate within the uncertainty of the data, but we cannot explain the value of $Z = 0.0134$ reported by \citet{asplund2009}. This suggests that the influence of accretion of metal-poor material from the protoplanetary disc onto the Sun cannot be ruled out.

The effect of PP disc accretion is, however, far more important for stars more massive than the Sun. The evolution of high-mass stars, M $> 2\Msun$, in particular, is characterised by very rapidly receding convective envelopes. During the first few million years of stellar evolution, the stars almost entirely lose their envelopes. Hence, accretion of matter on timescales typical of the PP disc lifetimes may significantly change the mass and surface metallicity of a star. The addition of matter and re-adjustment of the chemical composition in the convective envelope, in return, leads to changes in the evolutionary track of a star and, consequently, in the position of a star in the plane of observables (the colour-magnitude or $\teff$ - absolute magnitude diagrams). Our detailed calculations demonstrate that stars that have experienced metal-poor accretion move to the left, towards hotter $\teff$, and become more luminous. On the other hand, metal-rich accretion shifts a star to the right, towards cooler $\teff$ and lower luminosities. The effect ultimately depends on the initial metallicity and mass of a star, as well as the lifetime of the protoplanetary disc, but it would usually go unnoticed in composite stellar population or in a random sample of stars in a galaxy, because they represent a mixture of objects with different ages, metallicities, and accretion histories.

The effect of accretion can be best seen in the evolution of a coeval stellar population, such as an open cluster. Our detailed calculations show that the metal-rich and metal-poor accretion scenarios lead to a characteristic broadening of the cluster main-sequence (Fig. \ref{fig:Cartoon}). Given the stochastic nature of the disc formation problem, that is, the fact that different stars in a cluster will experience different accretion histories, it is natural to expect a dispersion of stars around some representative locus, which is canonically characterised by a "best-fit" isochrone. This dispersion is the key parameter that defines the properties of the disc accretion histories for individual stars in the cluster: it is largest at the main-sequence turn-off and it increases with decreasing the age of a cluster. Also protracted accretion histories, owing to, for example, longer disc lifetimes, may further increase the MSTO spread.

Remarkably, our model predictions for the MSTO dispersion appear to be in line with the observations of eMSTOs in open clusters of the Milky Way. To test the model, we employed six nearby Gaia DR2 clusters, which are well-populated on the MSTO and have robust measurements of kinematics, distance, and magnitudes for the individual members. The uncertainties of the data are extremely small, which rules out erroneous measurements as a viable explanation for the spread in the cluster CMD.
Indeed, all of these clusters do not only show the MSTO dispersion, but they also conform to the MSTO dispersion predicted by our simple coupled accretion - stellar evolution model. Younger clusters - Blanco 1, $\alpha$ Per, and Pleiades -  have very extended MSTOs, with dispersions exceeding $1000$ K. They are consistent with our model predictions for a disc with the lifetime of, at least, 3 Myr. It is also interesting that clusters older than $\sim 600$ Myr - Coma Ber, Praesepe,  Hyades - appear to have even larger MSTO spreads than expected from our model for their age. This may signify longer-lived discs.

In either case, precise astrometric and photometric observations by Gaia DR2 seem to support the scenario in which stars undergo a substantial change in their evolutionary tracks owing to the accretion from their natal protoplanetary discs. This intriguing signature can furthermore be explored with expanded samples of clusters with precision photometry, but also with detailed metallicity measurements of the cluster members. In particular, our model predicts a systematic variation of metallicity across the TO of a cluster. Stars that experienced the accretion of metal-poor material (gas) from their discs are shifted to the left, towards bluer colours (or higher $\teff$), whereas the objects that underwent the accretion of metal-rich material are displaced to the right, towards redder colours (or lower $\teff$). This peculiar effect, hence, might offer an indirect way to constrain the accretion histories and properties of protoplanetary discs of stars.

\textit{Acknowledgments.}
A.S.~is partially supported by grants ESP2017-82674-R (Spanish Government) and 2017-SGR-1131 (Generalitat de Catalunya). B.B. thanks the European Research Council (ERC Starting Grant 757448-PAMDORA) for their financial support. This study is supported by SFB 881 of the DFG (subprojects A05, A10).

\bibliographystyle{aa}
\bibliography{Citations}

\begin{appendix}
\section{Gaia photometry errors}

\begin{figure}[ht!]
  \centering
  \includegraphics[scale=0.9]{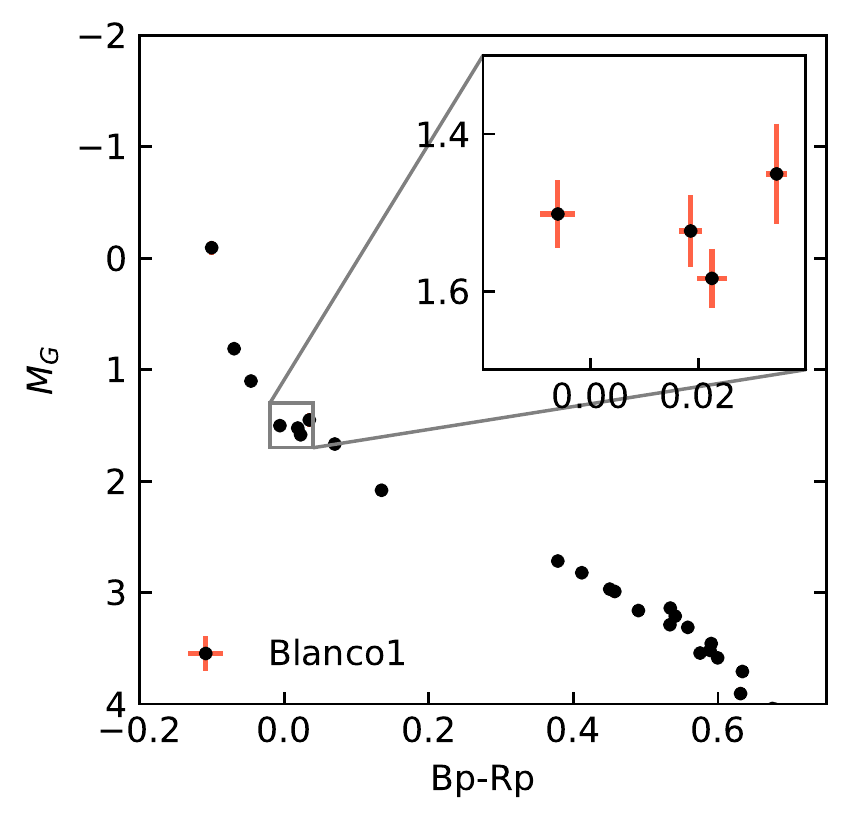}
  \caption{Gaia photometry error bars on Blanco1, with Blanco1 being the most distant cluster of our set of clusters. The errorbars are plotted for every single point but hidden by the markers in the zoomed-out view.}
  \label{fig:ErrBlanco}
\end{figure}

\begin{figure}[ht!]
  \centering
  \includegraphics[scale=0.9]{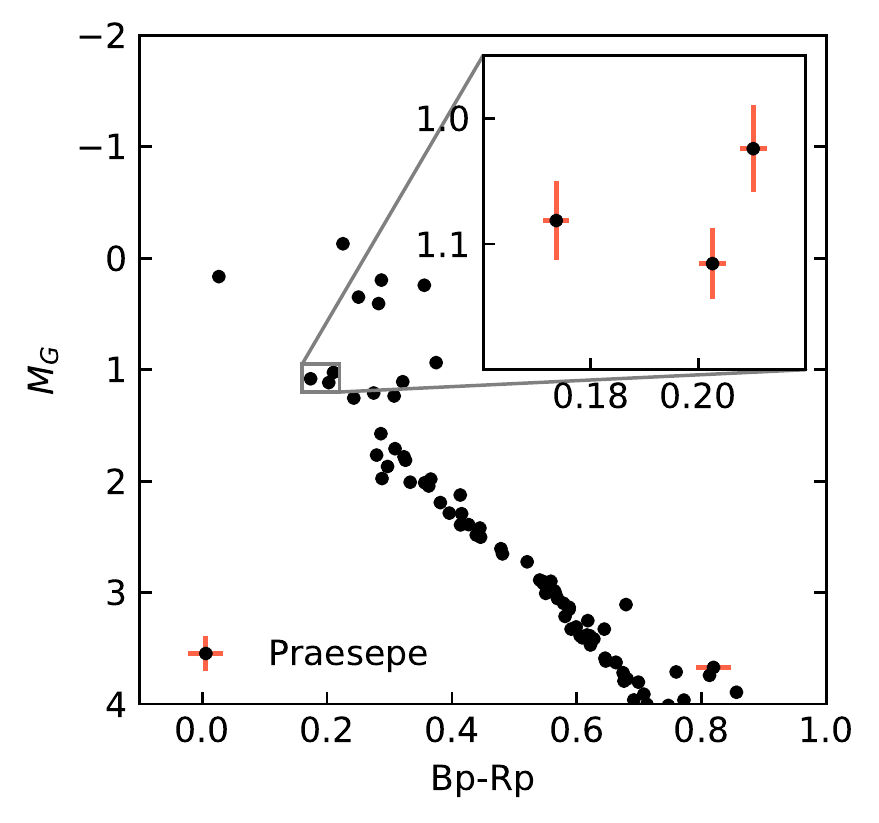}
  \caption{Gaia photometry error bars on Praesepe. The errorbars are plotted for every single point but hidden by the markers in the zoomed-out view for most stars.}
  \label{fig:ErrPraesepe}
\end{figure}

\end{appendix}

\end{document}